\documentstyle[aps,eqsecnum]{revtex}
\begin{document}

\def\psibar{\mbox{$\bar{\psi}$}}
\def\lambdabar{\mbox{$\bar{\lambda}$}}
\def\Htilde{\mbox{$\tilde{H}$}}
\def\Pbar{\mbox{$\bar{P}$}}
\def\Pbarhat{\mbox{$\hat{\bar{P}}$}}
\def\Phat{\mbox{${\hat P}$}}
\def\nablao{\stackrel{\circ}{\nabla}{\!\!}}
\def\go{\stackrel{\circ}{g}{\!\!}}
\def\Ro{\stackrel{\circ}{R}{\!\!}}
\def\Go{\stackrel{\circ}{G}{\!\!}}
\def\eo{\stackrel{\circ}{e}{\!\!}}
\def\calFo{\stackrel{\circ}{\cal F}{\!\!}}
\def\muhat{\hat{\mu}}
\def\nuhat{\hat{\nu}}
\def\alphahat{\hat{\alpha}}
\def\betahat{\hat{\beta}}
\def\rhohat{\hat{\rho}}
\def\sigmahat{\hat{\sigma}}

\title{Mass spectrum of ${\cal N}=8$ supergravity on $AdS_2 \times S^2$} 

\author{Steven Corley\thanks{scorley@phys.ualberta.ca}}

\address{Theoretical Physics Institute,
Department of Physics, University of Alberta,
Edmonton, Alberta, Canada T6G 2J1}

\maketitle
\begin{abstract}
An initial step is taken in investigating the duality
between the near horizon region of a four dimensional extremal
Reissner-Nordstr\"{o}m 
black hole and
the $n$-particle, ${\cal N}=4$ Calogero model as conjectured
by Gibbons and Townsend.  Specifically
we compute the mass spectrum of $d=4$, ${\cal N}=8$ supergravity about the
Bertotti-Robinson solution and find the corresponding set
of conformal dimensions of states in the dual conformal quantum
mechanics.  We find that the dual states fill irreducible representations
of the supergroup $SU(1,1|2)$, and furthermore transform
under various irreducible representations of the group $SU(2) \times
SU(6)$ spontaneously
broken down from the $E_{7(7)}$ duality group of ${\cal N}=8$ supergravity. 
\end{abstract}
\pacs{}

\section{Introduction}

The $AdS_2/CFT_1$ duality conjectured by Maldacena \cite{Mal} has  
received relativity little attention as compared to some of it's
higher dimensional cousins.  It is however one of the most interesting
cases with regard to black holes as the geometry $AdS_2
\times S^2$ arises as the near horizon geometry of a
Reissner-Nordstr\"{o}m black hole.  While the higher dimensional
$AdS/CFT$ dualities have been primarily used to learn about gauge
theory through gravity, one hopes that the reverse can be done
for four-dimensional black holes.  That is to say, it seems natural
that a conformal quantum mechanics will be simpler than a
supergravity theory on $AdS_2$.  Some recent investigations into
the $AdS_2/CFT$ duality are given in \cite{Str,GibTow}.

Since the $AdS/CFT$ duality was discovered by studying D-brane configurations,
in the hope of applying the duality to 4 dimensional 
black holes it is reasonable to
begin by looking for such solutions which have a D-brane interpretation.
One such solution \cite{D3soln} consists of four sets of D-branes in which 
any pair
intersects over a string.  Wrapping the D-branes on a six-torus and
taking each set of the four intersecting D-branes to have equal charge
gives rise to the extreme Reissner-Nordstr\"{o}m solution.  The near
horizon geometry of this solution is well known to be $AdS_2 \times S^2$,
with isometry supergroup $SU(1,1|2)$.
The results of \cite{KalRaj} further indicate that in fact $AdS_2 \times S^2$
is a solution to the full type IIB string theory and therefore that $SU(1,1|2)$
must also be a symmetry of the dual conformal quantum mechanics.
Gibbons and Townsend \cite{GibTow} have conjectured that the dual $CFT$
is given by the $n$-particle, ${\cal N} = 4$ superconformal Calogero 
model, which has yet to be constructed for arbitrary $n$ (see \cite{AkuKud}
for the $n=1$ case.)

As a first step toward investigating this conjecture we consider the 
$AdS_2$ side of the duality.  The low energy limit of type II(A or B) string
theory on $T^6$ is the ${\cal N}=8$ supergravity theory of 
Cremmer and Julia \cite{CremJul}.  Off-shell this theory has
an $SO(8)$ symmetry while on-shell the symmetry is enhanced to an
$E_{7(7)}$ duality.  We expand this theory about the above D3-brane
solution in the near horizon limit to linearized order in the fluctuations
to find the mass spectrum.  Using the prescription
of Gubser, Klebanov, and Polyakov \cite{GubKlePol} and
Witten \cite{Wit} we then extract the conformal weights of the
dual $CFT$ states and show that they lie in irreducible representations
of $SU(1,1|2)$, consistent with the fact that six of the supersymmetries
are broken in the near horizon limit of the D3-brane solution \cite{KalKum}.
Furthermore the $E_{7(7)}$ duality symmetry is broken to $SU(2) \times SU(6)$
\cite{Lar} and indeed 
the dual $CFT$ fields lie in irreducible representations
of this group as well. 

The procedure for extracting the masses is well known.  For example,
type IIB supergravity on $AdS_5 \times S^5$ was considered in
\cite{Kim,Gun} and 11 dimensional supergravity on $AdS_4 \times S^7$
and $AdS_7 \times S^4$ (among other spacetimes) in \cite{11d}.
More recently the mass spectrum of 6 dimensional, $N=4b$ supergravity
on $AdS_3 \times S^3$ \cite{Deg} was found for which the dual $CFT$ is
2 dimensional and detailed checks on the correctness of the
conjectured duality could be made, see \cite{review} and
references therein.  Other cases considered recently include
5 dimensional simple supergravity on $AdS_3 \times S^2$ \cite{Fuj1}
and $AdS_2 \times S^3$ \cite{Fuj2}, and ${\cal N}=8$ supergravity
on $AdS_3 \times S^2$ \cite{Lar}.  In this paper we follow closely
the procedure of \cite{Kim} to compute the mass spectrum
of 4 dimensional ${\cal N}=8$ supergravity about $AdS_2 \times S^2$.

In section \ref{section2} we briefly describe the $d=4$, ${\cal N} = 8$
supergravity theory of Cremmer and Julia \cite{CremJul} and 
give the Bertotti-Robinson (BR) solution \cite{BerRob} consisting of
an $AdS_2 \times S^2$ geometry with a nonvanishing two-form
flux on the two-sphere.
In section \ref{section3} we diagonalize the bosonic
equations of motion to obtain their mass spectrum, and repeat the
procedure in section \ref{section4} for the fermionic fields.  In
section \ref{section5} we go on to compute the conformal dimensions
of the corresponding states of the conformal quantum mechanics
model and demonstrate that they lie in irreducible representations
of $SU(1,1|2)$.  In section \ref{section6} we end with some conclusions.

We work in the signature $(-,+,+,+)$.

\section{${\cal N} = 8$ supergravity}
\label{section2}

\subsection{Field content}

The field content of $d=4$, ${\cal N} = 8$ supergravity consists
of a graviton described by a vierbein\footnote{Hatted
Greek indices are used for 4D coordinate indices whereas unhatted Greek
indices denote 2D coordinate indices with $\mu,\nu,...$=0,1 and
$\alpha, \beta, ...$=2,3.} $e_{\muhat}^{\;\;a}$, 8 gravitinos
described by the Majorana Rarita-Schwinger fields $\psi_{\muhat A}$
for $A=1,...,8$, 28 vector fields described by the Abelian gauge
fields $B_{\muhat}^{MN}$ (antisymmetric in $M,N$) for $M,N=1,...,8$,
56 spin-1/2 fields described by the Majorana spinors $\lambda_{ABC}$
(antisymmetric in $A,B,C$) for $A,B,C = 1,...,8$, and 70 spin-0
fields desribed by the scalar fields $W_{ABCD}$.  
The $O(8)$ invariant Lagrangian density describing this theory is given by
\begin{eqnarray}
{\cal L} & = & \biggl(\frac{1}{4} e R(\omega,e) + \frac{1}{2} 
\epsilon^{\muhat \nuhat \rhohat \sigmahat}
\psibar_{\muhat A} \gamma_{\sigmahat} \gamma_5 (\delta_{A}^{\;\;\;B}
D_{\nuhat}(\omega) - Q_{\nuhat A}^{\;\;\;\;\;B}) \psi_{\rhohat B} 
+ \frac{1}{8} e
G_{\muhat \nuhat}^{MN}(B) \Htilde_{MN}^{\muhat \nuhat}(B,{\cal V},\psi,\lambda)
\nonumber \\
& - & \frac{1}{12}i e \lambdabar_{ABC} \gamma^{\muhat} 
(\delta_{A}^{\;\;\;D} D_{\muhat}(\omega)
- 3 Q_{\muhat A}^{\;\;\;\;\;D}) \lambda_{BCD} 
-  \frac{1}{24} e P_{\muhat ABCD} \Pbar^{\muhat ABCD} -i \frac{1}{6 \sqrt{2}}
\psibar_{\muhat A} \gamma^{\nuhat} \gamma^{\muhat}
(\Pbar_{\nuhat}^{ABCD} + \Pbarhat_{\nuhat}^{ABCD}) \lambda_{BCD} \nonumber \\
& + & \frac{1}{8 \sqrt{2}} \Bigl(i \psibar_{\nuhat A} 
\gamma^{[ \nuhat }
{\hat{\cal F}}_{AB} \gamma^{\muhat ]} \psi_{\nuhat B}
 - \frac{1}{\sqrt{2}} i \psibar_{\muhat C}
{\hat{\cal F}}_{AB} \gamma^{\muhat} \lambda_{ABC} + \frac{i}{72} \eta \,
\epsilon^{ABCDEFGH} \lambdabar_{ABC} {\hat{\cal F}}_{DE} \lambda_{FGH}\Bigr)
\biggr)
\label{Lfull}
\end{eqnarray}
where
\begin{mathletters}
\begin{eqnarray}
G_{\muhat \nuhat}^{MN} & = & 2 \partial_{[\muhat} B^{MN}_{\nuhat]} \\
\Phat_{\muhat ABCD} & = & P_{\muhat ABCD} + 
i 2 \sqrt{2} \bigl( \psibar^{(L)}_{\muhat [A}
\lambda^{(R)}_{BCD]} + \frac{1}{24} \eta \, \epsilon_{ABCDEFGH} 
\psibar^{(R)E}_{\muhat} \lambda^{(L)FGH} \bigr) \\
\hat{\cal F}_{AB} & = & \gamma^{\muhat \nuhat} 
\hat{\cal F}_{AB \muhat \nuhat} \\
\hat{\cal F}_{AB \muhat \nuhat} & = & {\cal F}^{(F)}_{AB \muhat \nuhat} + 
\sqrt{2} \bigl(i
\psibar^{(R)}_{[\muhat [A} \psi^{(L)}_{\nuhat] B]} - i \frac{1}{\sqrt{2}}
\psibar^{(L)C}_{[\muhat} \gamma_{\nuhat]} \lambda^{(L)}_{ABC} + \frac{i}{288}
\eta \, \epsilon_{ABCDEFGH} \lambdabar^{CDE}_{(L)} \gamma_{\muhat \nuhat}
\lambda^{FGH}_{(R)} \bigr) \\
D_{\muhat}(\omega) \lambda & = & (\partial_{\muhat} + 
\frac{1}{4} \omega_{\muhat}^{ab}
\gamma_{ab}) \lambda.
\end{eqnarray}
\end{mathletters}
The $(R),(L)$ superscripts on the fermions denote
right and left-handed components defined by
$\lambda^{(R/L)} := (1/2)(1 \pm \gamma_5) \lambda$ ($\gamma$-matrix
conventions are described in the appendix).  
${\cal F}^{(F)}_{AB \muhat \nuhat}$ and $\tilde{H}^{\muhat \nuhat}_{MN}$ are
defined below.

The action possesses an $SU(8)$ gauge symmetry with gauge field
$Q_{\muhat A}^{\;\;\;\;\;B}$.  Capital Latin letters $A,B,C,...$
denote $SU(8)$ indices with lower(upper) indices transforming
in the 8($\bar{8}$) representation.  When an infinitesimal $SU(8)$
transformation $\Lambda_{A}^{\;\;B}= \Lambda^{\prime\;\;B}_A
+ i \Lambda^{\prime \prime \;\;B}_A$ (where $\Lambda^{\prime\;\;B}_A
= - \Lambda^{\prime\;\;A}_B$ and $\Lambda^{\prime \prime \;\;B}_A
= \Lambda^{\prime \prime \;\;A}_B$)
acts on a Majorana Fermi field
the $i$ is replaced by $i \gamma_5$ to preserve the Majorana condition.
This also explains why some contracted $SU(8)$ indices in the
action (\ref{Lfull}) are both in the lower position.

The scalars $P_{\muhat ABCD}$ and $SU(8)$ gauge fields 
$Q_{\muhat A}^{\;\;\;\;\;B}$
can be grouped together into an element of $E_{7(7)}$ in the fundamental
56 representation as
\begin{eqnarray}
\partial_{\muhat}{\cal V}{\cal V}^{-1} = \left( \begin{array}{cc}
\mbox{$Q_{\muhat [A}^{\;\;\;\;\;[C} \delta_{B]}^{\;\;\;D]}$} & 
\mbox{$P_{\muhat ABCD}$} \\
\mbox{${\bar P}_{\muhat}^{\;\;ABCD}$} & 
\mbox{${\bar Q}_{\muhat \;\;\;\;\;[C}^{\;\;\;[A} 
\delta^{B]}_{\;\;\;D]}$}
\end{array} \right) 
\label{Vconnection}
\end{eqnarray} 
where ${\bar Q}_{\muhat \;\;\;\;\;A}^{\;\;\;B} := 
(Q_{\muhat A}^{\;\;\;\;\;B})^*$,
$\Pbar_{\muhat}^{\;\;ABCD} := (P_{\muhat ABCD})^*$, and $P_{\muhat ABCD}$
satisfies the constraint
\begin{equation}
P_{\muhat ABCD} = \frac{1}{24} \eta \, \epsilon_{ABCDEFGH} 
\Pbar_{\muhat}^{\;\;EFGH}.
\label{scalarconstraint}
\end{equation}
The field ${\cal V}$ in this sense is analogous to the vierbein 
$e_{\muhat}^{\;\;a}$ and tranforms under $SU(8)$ on the left and
$E_{7(7)}$ on the right.

On shell the theory also possesses an $E_{7(7)}$ symmetry acting
on the right of $\cal{V}$ and on the left of the column
vector
\begin{eqnarray}
\left( \begin{array}{c} \mbox{${\cal F}^{(F)}_{\muhat \nuhat MN}$} \\
\mbox{$\bar{{\cal F}}^{(F)MN}_{\muhat \nuhat}$} \end{array} \right)
:= \frac{1}{\sqrt{2}} \left( \begin{array}{c}
\mbox{$G^{MN}_{\muhat \nuhat} + i H_{\muhat \nuhat MN} $} \\
\mbox{$G^{MN}_{\muhat \nuhat} - i H_{\muhat \nuhat MN} $} \end{array}
\right),
\label{E7F}
\end{eqnarray}
where the $M,N$ indices are $E_{7(7)}$ indices.
For the vector field the $E_{7(7)}$ transformation simply mixes
the equations of motion and Bianchi identities given by
\begin{eqnarray}
\nabla^{\muhat} (\tilde{\cal{F}}^{(F)MN}_{\muhat \nuhat} \mp 
\tilde{\bar{\cal{F}}}^{(F)MN}_{\muhat \nuhat}) = 0
\end{eqnarray}
respectively.  

To convert $E_{7(7)}$ indices to $SU(8)$ indices one acts with ${\cal V}$,
eg.,
\begin{eqnarray}
\left( \begin{array}{c}
\mbox{${\cal F}^{(F)}_{\muhat \nuhat AB}$} \\
\mbox{$\bar{{\cal F}}_{\muhat \nuhat}^{(F)AB}$} \end{array} \right)
:=
{\cal V}
\left( \begin{array}{c} \mbox{${\cal F}^{(F)}_{\muhat \nuhat MN}$} \\
\mbox{$\bar{{\cal F}}^{(F)MN}_{\muhat \nuhat}$} \end{array} \right).
\end{eqnarray}
$H_{\muhat \nuhat MN}$ is then eliminated from the Lagrangian
(\ref{Lfull}) by solving
\begin{eqnarray}
\hat{\cal F}_{\muhat \nuhat AB} = i \tilde{\hat{\cal F}}_{\muhat \nuhat AB}
\label{determineH}
\end{eqnarray}
where we define the dual by $\tilde{\cal F}^{\muhat \nuhat} := (1/2) 
e^{\muhat \nuhat \rhohat \sigmahat} {\cal F}_{\rhohat \sigmahat}$ where
$e_{\muhat \nuhat \rhohat \sigmahat}$ is the four-dimensional
volume element and we
follow the convention that $e_{0123}= - e$.  In the next
subsection we fix the $SU(8)$ symmetry and find expressions
for $\tilde{H}_{\muhat \nuhat MN}$ and $P_{\muhat ABCD}$.

Finally the theory is invariant under eight supersymmetries.
For the fermions the infinitesimal supersymmetry transformations
are given by
\begin{mathletters}
\begin{eqnarray}
\delta_S \lambda_{ABC} & = & \Bigl(\sqrt{2} \hat{P}_{\muhat ABCD} 
\gamma^{\muhat}
\epsilon^D - \frac{3}{4} \hat{\cal F}_{[AB} \epsilon_{C]} \Bigr)
\label{lambdasusy} \\
\delta_S \psi_{\muhat A} & = & \Bigl((\delta_{A}^{\;\;B} D_{\muhat}(\omega) -
Q_{\muhat A}^{\;\;\;\;\;B}) \epsilon_{B} + \frac{1}{4 \sqrt{2}}
\hat{\cal F}_{AB} \gamma_{\muhat} \epsilon_B + O(\bar{\lambda} \lambda,
\bar{\psi} \lambda) \epsilon \Bigr).
\label{psisusy}
\end{eqnarray}
\end{mathletters}
The supersymmetry transformations of the bosonic fields can
be found in \cite{CremJul}.

\subsection{Bertotti-Robinson background}

The generic 56 charge black hole solution to ${\cal N}=8$ supergravity
can be obtained by applying an $E_{7(7)}$ duality transformation to
a 5-parameter generating solution \cite{CveHul}.  In particular
an arbitrary 56 charge black hole with fixed moduli is obtained
by applying an arbitrary element of $SU(8)$($\subset E_{7(7)}$)
to the 5-parameter solution with the same moduli.  The 56 charge
black hole with arbitrary moduli is then obtained by applying
an $E_{7(7)}$ transformation.  The 56 ``dressed'' charges fit
into an $E_{7(7)}$ invariant antisymmetric matrix $Z_{AB}$
($A,B = 1, \cdots, 8$) transforming in the ${\bf 28}$ of
$SU(8)$.  This matrix can be block diagonalized via an $SU(8)$
transformation as
\begin{eqnarray}
Z_{AB} = \left( \begin{array}{cccccccc}
0 & \mbox{$\lambda_1$} & 0 & 0 & 0 & 0 & 0 & 0 \\
\mbox{$- \lambda_1$} & 0 & 0 & 0 & 0 & 0 & 0 & 0 \\
0 & 0 & 0 & \mbox{$\lambda_2$} & 0 & 0 & 0 & 0 \\
0 & 0 & \mbox{$-\lambda_2$} & 0 & 0 & 0 & 0 & 0 \\
0 & 0 & 0 & 0 & 0 & \mbox{$\lambda_3$} & 0 & 0 \\
0 & 0 & 0 & 0 & \mbox{$-\lambda_3$} & 0 & 0 & 0 \\
0 & 0 & 0 & 0 & 0 & 0 & 0 & \mbox{$-\lambda_4$} \\
0 & 0 & 0 & 0 & 0 & 0 & \mbox{$-\lambda_4$} & 0 \\
\end{array} \right).
\label{ccmatrix}
\end{eqnarray}
The unbroken global symmetry group of the solution is then given
by the subgroup of $SU(8)$ that leaves $Z_{AB}$ invariant.
A particular case of interest are the extreme black holes, which
can be put in the form (\ref{ccmatrix})
with $\lambda_2=\lambda_3=\lambda_4=0$, and therefore have
$SU(2) \times SU(6)$ global symmetry group \cite{Lar}.  It follows 
that any extreme
black hole can be obtained from any other extreme black hole by an
$SU(8)$ transformation (and then an $E_{7(7)}$ transformation to
change the moduli if necessary).  Since the mass spectrum is invariant
under these transformations then knowing it for one means that we
know it for all, and in particular for the intersecting D3-brane
solution of \cite{D3soln}.  We are actually only interested in the near
horizon limit, but clearly the same argument applies.  Rather than
expanding about the near horizon limit of the intersecting D3-brane
solution, which is $AdS_2 \times S^2$ with four non-vanishing
vector fields, we instead expand about the same geometry but
with only one non-zero vector with flux on $S^2$ only.  We now
proceed to describe this background.

We consider the compactification of the ${\cal N}=8$ supergravity
theory (\ref{Lfull}) about the BR solution consisting of an
$AdS_2 \times S^2$ geometry with curvature tensors
\begin{eqnarray}
\Ro_{\mu \nu \lambda \rho} & = & -\frac{1}{l^2}(\go_{\mu \lambda} \go_{\nu \rho}
- \go_{\mu \rho} \go_{\nu \lambda}) \\
\Ro_{\alpha \beta \gamma \delta} & = & \frac{1}{l^2}(\go_{\alpha \gamma} 
\go_{\beta \delta}
- \go_{\alpha \delta} \go_{\beta \gamma}) 
\label{curvatures}
\end{eqnarray}
where $\mu,\nu,...$ are curved $AdS_2$ indices,
$\alpha,\beta,...$ 
curved $S^2$ indices, and $l$ is the radius of curvature of both spaces.
The only other non-vanishing field that we consider is a two-form flux on
$S^2$ which we take to be\footnote{We put square brackets around the 
$M,N$ indices when they take specific numerical values in order
to avoid confusion with other indices.}
\begin{eqnarray}
\Go^{[12]}_{\mu \nu} = 0, \hspace{1cm} \Go^{[12]}_{\alpha \beta} = \frac{1}{l},
e_{\alpha \beta}.
\label{Gbackground}
\end{eqnarray}
i.e., the Freund-Rubin ansatz \cite{Fre}.
The $AdS_2$ and $S^2$ volume
elements are denoted $e_{\mu \nu}$ and $e_{\alpha \beta}$ respectively
and are related to the four-dimensional volume element by
$e_{\mu \nu \alpha \beta}= - e_{\mu \nu} e_{\alpha \beta}$.
The remaining fields in (\ref{Lfull}) vanish except for the
vielbein ${\cal V}$ which is equal to the identity
\begin{eqnarray}
{\cal V} = \left( \begin{array}{cc}
\delta_{[A}^{\;\;[M} \delta_{B]}^{\;\;N]} & 0 \\
0 & \delta^{[A}_{\;\;[M} \delta^{B]}_{\;\;N]} 
\end{array} \right).
\label{Vback}
\end{eqnarray}

That these fields actually solve the equations of motion is
clear from the expression (\ref{Hbosonic})
for $\tilde{H}_{\muhat \nuhat MN}$ given in the next section.  When
all fields vanish except for $g_{\muhat \nuhat}$ and $B_{\muhat}^{[12]}$
and ${\cal V}$ is given by (\ref{Vback}) then
\begin{equation}
\tilde{H}^{MN}_{\muhat \nuhat} = 
-\delta_{[1}^{\;\;[M} \delta_{2]}^{\;\;N]}
G^{[12]}_{\muhat \nuhat},
\end{equation}
in which case the action (\ref{Lfull}) reduces to the 
Einstein-Maxwell action.

The BR solution preserves only two of the eight supersymmetries
\cite{KalKum}.  To see this set the fermionic supersymmetry
transformations (\ref{lambdasusy},\ref{psisusy}) to zero.  It
immediately follows from $\delta_S \lambda_{ABC} =0$ that
$\epsilon^C=0$ for $C \neq 1,2$.  The remaining conditions are
satisfied trivially except for $\delta_S \psi_{\hat{\mu} A} = 0$
for $A=1,2$ which implies the Killing spinor equations
\begin{eqnarray}
D_{\mu}(\omega) \tilde{\eta}_A & - & \frac{i}{2l} \rho_{\mu} \epsilon_{AB}
\tilde{\eta}_B = 0 \\
D_{\alpha}(\omega) \eta_A & - & \frac{i}{2l} \rho_5 \rho_{\alpha} \epsilon_{AB}
\eta_B = 0 
\end{eqnarray}
where we have decomposed $\epsilon^A$ into a product of
two-component spinors as $\epsilon^A = \tilde{\eta}^A \otimes \eta^A$
and similarly for the $\gamma$-matrices as discussed in the appendix.

\subsection{Linearized equations of motion}

To find the equations of motion we must solve (\ref{determineH})
for $\tilde{H}_{\muhat \nuhat MN}$.  To do this 
it is convenient to fix the
$SU(8)$ gauge symmetry to the so called symmetric gauge where
${\cal V} = \exp(X)$ and
\begin{eqnarray}
X = \left( \begin{array}{cc}
0 & \mbox{$W_{ABCD}$} \\
\mbox{$\bar{W}^{ABCD}$} & 0 \end{array} \right).
\label{gauge}
\end{eqnarray}
$W_{ABCD}$ is complex, completely antisymmetric in $A,B,C,D$, and satisfies
the constraint
\begin{eqnarray}
\bar{W}^{ABCD} = \frac{1}{24} \eta \, \epsilon^{ABCDEFGH} W_{EFGH}.
\label{Wconstraint}
\end{eqnarray}
In this gauge it is straightforward to determine $\tilde{H}_{\muhat \nuhat MN}$
from (\ref{determineH}).  Since our interest is in computing the mass
spectrum of the theory about the BR background we only need
$\tilde{H}_{\muhat \nuhat MN}$ to quadratic order in fluctuations, therefore
it is sufficient to consider
\begin{eqnarray}
G^{MN}_{\muhat \nuhat} \tilde{H}^{(B)}_{\muhat \nuhat MN} = 
-G^{MN}_{\muhat \nuhat} (1+W+\bar{W}+W^2+\bar{W}^2)_{MNPQ} 
G^{\muhat \nuhat PQ} + i G^{MN}_{\muhat \nuhat}(W-\bar{W}+W^2-\bar{W}^2)_{MNPQ} 
\tilde{G}^{\muhat \nuhat PQ}
\label{Hbosonic}
\end{eqnarray}
for the purely bosonic part of $\tilde{H}_{\muhat \nuhat MN}$, and
\begin{eqnarray}
G^{MN}_{\muhat \nuhat} \tilde{H}^{(F) \muhat \nuhat}_{MN} = 
\frac{1}{\sqrt{2}} \bigl(i \psibar_{\nuhat A} \gamma^{[ \nuhat }
\calFo_{AB} \gamma^{\muhat ]} \psi_{\nuhat B} - \frac{1}{\sqrt{2}} 
i \psibar_{\muhat C}
\calFo_{AB} \gamma^{\muhat} \lambda_{ABC} + \frac{i}{72} \eta \,
\epsilon^{ABCDEFGH} \lambdabar_{ABC} \calFo_{DE} \lambda_{FGH}\bigr)
\label{Hfermionic}
\end{eqnarray}
for the fermionic part, where 
\begin{eqnarray}
\stackrel{\circ}{\cal F}_{AB} := 
\frac{1}{\sqrt{2}} (G_{\muhat \nuhat AB} + i \gamma_5
\tilde{G}_{\muhat \nuhat AB})\gamma^{\muhat \nuhat}.
\end{eqnarray}
The solution to (\ref{determineH}) is not unique.  We have used this
freedom to simplify the equations of motion as much as possible.

Expressions for $Q_{\muhat A}^{\;\;\;\;\;B}$ and $P_{\muhat ABCD}$ follow 
straightforwardly in this gauge from (\ref{Vconnection}) and
(\ref{gauge}).  
$Q_{\muhat A}^{\;\;\;\;\;B}$ is at least linear in $W_{ABCD}$ and therefore
can be dropped in the action as it always multiplies a pair
of fermions.  $P_{\muhat ABCD}$ however is given by
\begin{equation}
P_{\muhat ABCD} = \partial_{\muhat} W_{ABCD}
\label{PinW}
\end{equation}
and contributes to the quadratic part of the action.

To obtain the linearized equations of motion, substitute 
(\ref{Hbosonic},\ref{Hfermionic},\ref{PinW}) into the
action (\ref{Lfull}), expand $g_{\muhat \nuhat}$ and 
$G^{MN}_{\muhat \nuhat}$ about
the background (\ref{curvatures},\ref{Gbackground}) as
\begin{mathletters}
\begin{eqnarray}
g_{\muhat \nuhat} & = & \go_{\muhat \nuhat} + h_{\muhat \nuhat} \\
G^{MN}_{\muhat \nuhat} & = & \Go_{\muhat \nuhat}^{MN} + g^{MN}_{\muhat \nuhat},
\end{eqnarray}
\end{mathletters}
vary, and keep only linear terms in the fluctuations in the resulting
equations of motion.
For the bosonic fields we find
\begin{eqnarray}
\Bigl( & - & \frac{1}{2} \nabla^2 h_{\muhat \nuhat} - \frac{1}{2}
\nabla_{\muhat} \nabla_{\nuhat}h + 
\nabla_{(\muhat}\nabla^{\alphahat}
h_{\nuhat) \alphahat} + \Ro_{(\muhat}^{\alphahat} h_{\nuhat) \alphahat}
+ \Ro_{\alphahat \muhat \nuhat \betahat} h^{\alphahat \betahat}
-\frac{1}{2} h_{\muhat \nuhat} \Ro 
- \frac{1}{2} \go_{\muhat \nuhat}
\bigl( - \nabla^2 h + \nabla_{\alphahat} \nabla_{\betahat}
h^{\alphahat \betahat} - h^{\alphahat \betahat} \Ro_{\alphahat \betahat}
\bigr) \Bigr) \nonumber \\
& = & 2 \Bigl( 2 \Go_{(\muhat}^{[12]\alphahat} 
g^{[12]}_{\nuhat) \alphahat} - \Go^{[12]}_{\muhat \alphahat} 
\Go^{[12]}_{\nuhat \betahat} h^{\alphahat \betahat} - \frac{1}{2}
\go_{\muhat \nuhat} \bigl(\Go^{[12] \alphahat \betahat} 
g^{[12]}_{\alphahat \betahat} - \Go_{\;\;\;\;\alphahat}^{[12]\;\;\nuhat}
\Go^{[12]}_{\betahat \nuhat} h^{\alphahat \betahat} \bigr)
- \frac{1}{4} h_{\muhat \nuhat} \Go^{[12] \alphahat \betahat}
\Go^{[12]}_{\alphahat \betahat} \Bigr)
\label{Einstein}
\end{eqnarray}
for the linearized Einstein equations,
\begin{mathletters}
\begin{eqnarray}
\nabla_{\muhat} \bigl(g^{[12] \muhat \nuhat} - 2 
\Go^{[12][\muhat}_{\;\;\;\;\alphahat} h^{\nuhat] \alphahat}
+\frac{1}{2} \Go^{[12] \muhat \nuhat} h \bigr) & = & 0 \hspace{1cm} M=1,N=2 
\label{V1} \\
\nabla_{\muhat} g^{[MN] \muhat \nuhat} & = & 0 \hspace{1cm} M=1,2;N\neq 1,2
\label{V6} \\
\nabla_{\muhat} \bigl(g^{[MN] \muhat \nuhat} + 2 \Go^{[12]\muhat \nuhat}
(W + \bar{W})_{12MN} - i 2 \widetilde{\Go}{\!\,}^{[12] \muhat \nuhat} (W
-\bar{W})_{12MN} \bigr) & = & 0 \hspace{1cm} M,N \neq 1,2
\label{V15}
\end{eqnarray}
\end{mathletters}
for the linearized vector equations, and
\begin{mathletters}
\begin{eqnarray}
\bigl(\nabla^2 W_{12CD} - (\Go^{[12] \muhat \nuhat} + i
\widetilde{\Go}{\!\,}^{[12] \muhat \nuhat}) g_{\muhat \nuhat}^{CD}
- \Go^{[12]}_{\muhat \nuhat} (\Go^{[12] \muhat \nuhat} + i
\widetilde{\Go}{\!\,}^{[12] \muhat \nuhat}) \bar{W}^{12CD} \bigr) =  
0 \hspace{0.5cm} C,D & \neq & 1,2 \label{S15} \\
\nabla^{2} W_{ABCD} = 0 \hspace{0.5cm} A=1,2;B,C,D & \neq & 1,2 
\label{S20} \\
\end{eqnarray}
\end{mathletters}
for the linearized scalar equations.  All remaining bosonic equations
can be obtained from the above equations by symmetry, complex
conjugation, and the 
constraint (\ref{Wconstraint}).

The linearized fermion equations of motion follow in a similar way.
We find
\begin{mathletters}
\begin{eqnarray}
\bigl( \gamma^{\muhat} D_{\muhat} \lambda_{12C} + \frac{1}{4}
\gamma^{\muhat} \calFo_{[12]} \psi_{\muhat C} \bigr) & = & 0
\hspace{1cm} C \neq 1,2 \label{Sp6} \\
\gamma^{\muhat} D_{\muhat} \lambda_{ABC} 
& = & 0
\hspace{1cm} A=1,2; B,C \neq 1,2 \label{Sp15} \\
\bigl( \gamma^{\muhat} D_{\muhat} \lambda_{ABC} - \frac{\eta}{12 \sqrt{2}}
\epsilon^{12ABCFGH} \calFo_{[12]} \lambda_{FGH} \bigr) & = & 0
\hspace{1cm} A,B,C \neq 1,2 \label{Sp20} \\
\end{eqnarray}
\end{mathletters}
for the spinors and
\begin{mathletters}
\begin{eqnarray}
\bigl( e^{\muhat \nuhat \rhohat \sigmahat} 
\gamma_{\sigmahat} \gamma_5 D_{\nuhat}
\psi_{\rhohat A} + \frac{i}{2 \sqrt{2}} \gamma^{[ \nuhat} \calFo_{AB}
\gamma^{\muhat]} \psi_{\nuhat B} \bigr) & = & 0 \hspace{1cm} A=1,2 
\label{G1} \\
\bigl(e^{\muhat \nuhat \rhohat \sigmahat} \gamma_{\sigmahat} 
\gamma_5 D_{\nuhat}
\psi_{\rhohat A} - \frac{i}{4} \calFo_{[12]}
\gamma^{\muhat} \lambda_{12A} \bigr) & = & 0 \hspace{1cm} A \neq 1,2 
\label{G6}
\end{eqnarray}
\end{mathletters}
for the gravitinos.

\section{Bosonic masses}
\label{section3}
\subsection{Harmonic expansion on $S^2$}

To find the bosonic mass spectrum on $AdS_2$ we expand the fields in spherical
harmonics on $S^2$.  The expansions are quite simple in this case
as all harmonic functions on the 2-sphere can be expressed in terms
of just the scalar spherical harmonics $Y_{lm}$.  The expansions
of the bosonic fluctuations are then given by (denoting the $l,m$
indices collectively by $(k)$)
\begin{mathletters}
\begin{eqnarray}
h_{\mu \nu} & = & \sum_k H^{(k)}_{\mu \nu} Y_{(k)} \label{hmunuexp} \\
h_{\mu \alpha} & = & \sum_k (B^{(k)}_{1 \mu} \nabla_{\alpha} Y_{(k)}
+ B^{(k)}_{2 \mu} e_{\alpha \beta} \nabla^{\beta} Y_{(k)})
\label{hmualphaexp} \\
h_{\alpha \beta} & = & \sum_{k} (\phi^{(k)}_1 \nabla_{\alpha} \nabla_{\beta}
Y_{(k)} + \phi^{(k)}_2 e_{(\alpha}^{\;\;\;\gamma} \nabla_{\beta)}
\nabla_{\gamma} Y_{(k)} + \phi^{(k)}_3 g_{\alpha \beta} Y_{(k)})
\label{halphabetaexp} \\
b_{\mu}^{AB} & = & \sum_k b_{\mu}^{(k)AB} Y_{(k)} \label{bmuexp} \\
b_{\alpha}^{AB} 
& = & \sum_k (b^{(k)AB}_{1} \nabla_{\alpha} Y_{(k)}
+ b^{(k)AB}_{2} e_{\alpha \beta} \nabla^{\beta} Y_{(k)}) 
\label{balphaexp} \\
W_{ABCD} & = & \sum_k W^{(k)}_{ABCD} Y_{(k)},
\label{Wexp}
\end{eqnarray}
\end{mathletters}
where the harmonics satisfy\footnote{For the remainder of the paper
we set the curvature scale $l=1$ and drop the $\circ$ symbol above
background quantities.} $\nabla_{\alpha} \nabla^{\alpha} Y_{(k)}
= -k(k+1) Y_{(k)}$.

Before substituting the expansions into the linearized equations of
motion we can first simplify the expansions 
by fixing some of the gauge symmetries.
Specifically we have four dimensional diffeomorphism invariance and 
28 $U(1)$ gauge
invariances.  To fix the diffeomorphism invariance we work
in de Donder-Lorentz gauge
\begin{equation}
\nabla^{\alpha} h_{\alpha \mu} = 0 = \nabla^{\alpha}(h_{\alpha \beta}
- \frac{1}{2} g_{\alpha \beta} g^{\gamma \delta} h_{\gamma \delta}).
\end{equation}
The $U(1)$ invariances are fixed by the Lorentz-like gauge conditions
\begin{equation}
\nabla^{\alpha} b_{\alpha}^{AB} = 0.
\end{equation}
Plugging in the above expansions yields the conditions
\begin{eqnarray}
\phi_{1}^{(k)} = \phi_{2}^{(k)} = B_{1 \mu}^{(k)} & = & 0 \hspace{1cm} k>1 \\
b_{1}^{(k)AB} & = & 0 \hspace{1cm} k \geq 1. 
\label{diffeofix}
\end{eqnarray}

The gauge fixing conditions do not quite fix all the gauge symmetry, but rather
leave diffeomorphism and $U(1)$ gauge symmetries of the zero modes,
and further the conformal diffeomorphisms generated by
the vectors 
\begin{equation}
\xi_{\mu} = - \nabla_{\mu} \xi_{1}^{(1)} Y_{(1)}, \,
\xi_{\alpha} = ( \xi_{1}^{(1)} \nabla_{\alpha} Y_{(1)} + \xi_{2}^{(1)}
\eo_{\alpha \beta} \nabla^{\beta} Y_{(1)}).
\end{equation}
The zero mode symmetries we will treat later.  The conformal diffeomorphisms
are easily dealt with by noting that the $Y_{(1)}$ harmonics
satisfy 
\begin{equation}
(\nabla_{\alpha} \nabla_{\beta} + g_{\alpha \beta}) Y_{(1)} = 0.
\label{conformaldiffeos}
\end{equation}
It follows that the expansion of $h_{\alpha \beta}$ in the $Y_{(1)}$
sector contains only $\phi^{(1)}_3$ (after a redefinition).  Under
a conformal diffeomorphism $\phi^{(1)}_3$ transforms as
\begin{equation}
\delta \phi^{(1)}_3 = -2 \xi^{(1)}_1
\end{equation}
and therefore can be set to zero.  Similarly under a conformal
diffeomorphism generated by $\xi^{(1)}_2$ we find
\begin{equation}
\delta B^{(1)}_{2 \mu} = \nabla_{\mu} \xi^{(1)}_2.
\label{B12symm}
\end{equation}
To fix this symmetry we demand the Lorentz condition
\begin{equation}
\nabla^{\mu} B_{2 \mu}^{(1)} = 0.
\label{confdiffeogauge}
\end{equation}

\subsection{$AdS_2$ linearized equations of motion}

To keep things as simple as possible, we begin by considering
only the ${\cal N}=2$ supergravity fluctuations, $h_{\muhat \nuhat}$
and $b_{\muhat}^{[12]}$.  In terms of the global symmetry group
$SU(2) \times SU(6)$ both transform as singlets.  Substituting the 
expansions 
(\ref{hmunuexp}-\ref{balphaexp})
along with the background field strength (\ref{Gbackground}) 
into the linearized equations
of motion yields\footnote{We use the notation $\nabla^{2}_x := 
g^{\mu \nu} \nabla_{\mu} \nabla_{\nu}$.}
\begin{mathletters}
\begin{eqnarray}
\Bigl( \bigl(\nabla^{2}_x +2-k(k+1) \bigr) 
H^{(k)}_{\mu \nu} -2 \nabla_{(\mu}
\nabla^{\lambda} H^{(k)}_{\nu) \lambda} + \bigl(\nabla_{\mu} \nabla_{\nu}
- g_{\mu \nu} (\nabla^{2}_x +1 & - & k(k+1)) \bigr) H^{(k)} + g_{\mu \nu}
\nabla^{\lambda} \nabla^{\rho} H^{(k)}_{\lambda \rho} \nonumber \\
+ 2 \bigl(
\nabla_{\mu} \nabla_{\nu} - g_{\mu \nu} (\nabla^{2}_x -1-\frac{1}{2}k(k+1))
\bigr) \phi^{(k)}_3 \Bigr) & = & 4 g_{\mu \nu} k(k+1) b^{(k)}_2
\label{Emunu} \\
\Bigl( \bigl(\nabla^{2}_x -1 -k(k+1) \bigr) B^{(k)}_{2 \mu} -
\nabla_{\mu} \nabla^{\nu} B^{(k)}_{2 \nu} \Bigr) & = & 4 b_{\mu}^{(k)[12]}
\label{Emualpha1} \\
\bigl( \nabla^{\nu} H^{(k)}_{\mu \nu} - \nabla_{\mu} H^{(k)} - \nabla_{\mu}
\phi^{(k)}_3 \bigr) & = & -4 \nabla_{\mu} b^{(k)}_{2} 
\label{Emualpha2} \\
\Bigl( \bigl( \nabla^{2}_x + 4 \bigr) \phi^{(k)}_3 + 
\bigl(\nabla^{2}_x -1-k(k+1) \bigr)
H^{(k)} - \nabla^{\mu} \nabla^{\nu} H^{(k)}_{\mu \nu} \Bigr) & = & 
4 k(k+1) b^{(k)}_2 \label{Ealphabeta1} \\
H^{(k)} & = & 0 \label{Ealphabeta2} \\
\nabla^{\mu} B^{(k)}_{2 \mu} & = & 0 \label{Ealphabeta3}
\end{eqnarray}
\end{mathletters}
for the Einstein equations and
\begin{mathletters}
\begin{eqnarray}
\bigl( 2\nabla^{\mu} \nabla_{[\mu} b_{\nu]}^{(k)[12]} - k(k+1)
b_{\nu}^{(k)[12]} - k(k+1) B^{(k)}_{2 \nu} \bigr) & = & 0 \label{V12mu} \\
\bigl( (\nabla^{2}_x - k(k+1)) b_{2}^{(k)} + \phi_{3}^{(k)} - \frac{1}{2}
H^{(k)} \bigr) & = & 0 \label{V12alpha1} \\
\bigl( \nabla^{\mu} B^{(k)}_{2 \mu} + \nabla^{\mu} b_{\mu}^{(k)[12]}
\bigr) & = & 0 \label{V12alpha2}
\end{eqnarray}
\end{mathletters}
for the $M=1,N=2$ vector equations.  All equations are valid for $k \geq 2$.
The $k=0,1$ cases will be handled separately.  Actually one notices that there
are more equations than fields.  An important consistency check is that the
equations are not all independent, as follows easily by taking divergences
on the $AdS_2$ index of the above equations.  More generally this follows
from the Bianchi identities.

The Einstein equations (\ref{Emunu}), (\ref{Emualpha2}) and 
(\ref{Ealphabeta2}) allow
one to eliminate the metric fluctuation $H^{(k)}_{\mu \nu}$ (locally)
in terms of the remaining fields.  This follows by a simple counting
argument\footnote{This argument has also been given recently in 
\cite{Mic}.}.  A homogeneous solution to these equations would have to
satisfy $H^{(k)}=0$ and $\nabla^{\mu} H^{(k)}_{\mu \nu}=0$.  These equations
are enough to determine $H^{(k)}_{\mu \nu}$ though, i.e., three equations
and three unknowns.  However $H^{(k)}_{\mu \nu}$ must also satisfy
the homogeneous part of (\ref{Emunu}), which clearly cannot be the
case in general for arbitrary $k$.  In fact one can show easily
that the traceless and divergenceless conditions imply
$(\nabla^{2}_x + 2)H^{(k)}_{\mu \nu}=0$, which is only consistent
with (\ref{Emunu}) when $k=0$.

The divergence equations
(\ref{Ealphabeta3}) and (\ref{V12alpha2}) remove one degree of freedom
each from the vectors, reducing them effectively to scalars
$(e^{\lambda \rho}\nabla_{\lambda} B^{(k)}_{2 \rho})$ and
$(e^{\lambda \rho}\nabla_{\lambda} b^{(k)[12]}_{\rho})$ respectively.
Acting on the vector equations (\ref{Emualpha1}) and (\ref{V12mu})
with the operator $e^{\lambda \mu} \nabla_{\lambda}$ and 
using the two-dimensional identity
\begin{equation}
2 \nabla_{[\mu} a_{\nu]} = - e_{\mu \nu} e^{\lambda \rho} \nabla_{\lambda}
a_{\rho} 
\label{2dident}
\end{equation}
we obtain the coupled scalar equations
\begin{mathletters}
\begin{eqnarray}
\bigl( (\nabla^{2}_x - k(k+1)) e^{\mu \nu} \nabla_{\mu} b_{\nu}^{(k)[12]}
- k(k+1) e^{\mu \nu} \nabla_{\mu} B^{(k)}_{2 \nu} \bigr) & = & 0 \\
\bigl( (\nabla^{2}_x - k(k+1)-2) e^{\mu \nu} \nabla_{\mu} B_{2 \nu}^{(k)}
- 4 e^{\mu \nu} \nabla_{\mu} b^{(k)[12]}_{\nu} \bigr) & = & 0.
\label{12vectors}
\end{eqnarray}
\end{mathletters}
Similarly substituting the Einstein equations (\ref{Emualpha2}) and
(\ref{Ealphabeta2}) into the $\phi^{(k)}_3, b_{2}^{(k)}$ equations
of motion (\ref{Ealphabeta1}) and (\ref{V12alpha1}) we obtain
the coupled equations
\begin{mathletters}
\begin{eqnarray}
\bigl( (\nabla^{2}_x - k(k+1)-2) \phi^{(k)}_3
+4 k(k+1) b^{(k)}_{2} \bigr) & = & 0 \\
\bigl( (\nabla^{2}_x - k(k+1)) b_{2}^{(k)}
+ \phi_{3}^{(k)} \bigr) & = & 0.
\label{12scalars}
\end{eqnarray}
\end{mathletters}
Defining the scalar mass by $(\nabla^{2}_x - m^2) \phi =0$, the respective
mass matrices are easily diagonalized and we find
\begin{equation}
m^2 = k(k-1), (k^2 + 3k + 2)
\label{12scalarmasses}
\end{equation}
in each case, where each mass at level $k \geq 2$ is $(2k+1)$ 
degenerate.
\vspace{5mm}

\noindent
$\bullet$ {\bf $k=1$ sector}
\vspace{5mm}

The $k=1$ sector equations of motion follow from the $k \geq 2$ equations
after taking into account the gauge fixing condition $\phi_{3}^{(1)}=0$ and
the relations (\ref{conformaldiffeos}).  The latter equations imply that
the Einstein equations (\ref{Ealphabeta1}) and (\ref{Ealphabeta2}) are 
not separate equations but rather are replaced by 
\begin{equation}
\bigl( (\nabla^{2}_x -2)H^{(1)} - \nabla^{\mu} \nabla^{\nu} H^{(1)}_{\mu \nu}
\bigr) = 8 b^{(1)}_2.
\end{equation}
Furthermore (\ref{Ealphabeta3}) no longer follows from the Einstein
equations but nevertheless holds due to our choice of gauge 
(\ref{confdiffeogauge}).  As a consequence the coupled equations
(\ref{12vectors}) for $(e^{\mu \nu} \nabla_{\mu} B_{2 \nu}^{(1)})$ and
$(e^{\mu \nu} \nabla_{\mu} b_{\nu}^{(1)[12]})$ continue to hold for
$k=1$, and therefore also the masses (\ref{12scalarmasses}).  However
at $k=1$ the mass $m^2 = k(k-1)$ vanishes and therefore one of the 
vectors must be massless, in particular one can
show that
\begin{equation}
(\nabla^{2}_x + 1) (B^{(1)}_{2 \mu} - \frac{2}{\sqrt{2}} b^{(1)[12]}_{\mu})
= 0.
\end{equation}
This field can be gauged away locally due to the residual gauge
transformations (\ref{B12symm}) 
of $B^{(1)}_{2 \mu}$, i.e., $\delta B^{(1)}_{2 \mu}
= \nabla_{\mu} \xi^{(1)}_2$ with $\nabla^{2}_x \xi^{(1)}_2 = 0$ implies
\begin{equation}
(\nabla^{2}_x + 1 ) \nabla_{\mu} \xi^{(1)}_2 = 0.
\end{equation}
Although this field can be removed locally, it still has boundary
degrees of freedom and will be important in filling a representation
of $SU(1,1|2)$ discussed later, so we shall continue to treat
it as a $k=1$ field.

By the same argument as in the $k \geq 2$ case above, 
$H^{(1)}_{\mu \nu}$ can be eliminated
in terms of $b_{2}^{(1)}$.  Specifically the trace of (\ref{Emunu}) implies
that $H^{(1)} = 8 b_{2}^{(1)}$, which along with (\ref{Emunu},\ref{Emualpha2})
is enough to eliminate $H^{(1)}_{\mu \nu}$.  Substituting for
$H^{(1)}$ in (\ref{V12alpha1}) results in
\begin{equation}
(\nabla^{2}_x - 6) b_{2}^{(1)} = 0.
\end{equation}
It follows that the mass spectrum (\ref{12scalarmasses}) holds for
$k=1$ for the 
scalars $b_{2}^{(1)}$ and $\phi_{3}^{(1)}$ for the mass
$m^2 = (k^2 + 3k + 2)$, but not for $m^2 = k(k-1)$.
\vspace{5mm}

\noindent
$\bullet$ {\bf $k=0$ sector}
\vspace{5mm}

In the zero mode sector $k=0$ the equations of motion are given by
(\ref{Emunu}), (\ref{Ealphabeta1}), and (\ref{V12mu}).  In this sector
though we still have two-dimensional diffeomorphism invariance and a
$U(1)$ gauge symmetry.  Before fixing these symmetries note that
the trace of (\ref{Emunu}) implies
\begin{equation}
(\nabla^{2}_x -2) \phi^{(0)}_3 = 0
\end{equation}
completing one of the $m^2=(k^2 + 3k + 2)$ towers above to $k=0$.

The remaining fields can be gauged away locally.  This can be seen by
first fixing the diffeomorphism symmetry by demanding
\begin{equation}
\nabla^{\mu} H^{(k)}_{\mu \nu} = 5 \nabla_{\nu} \phi^{(0)}_3.
\end{equation}
This implies from (\ref{Ealphabeta1}) that
\begin{equation}
(\nabla^{2}_x -1) (\phi^{(0)}_3 - \frac{1}{4} H^{(0)} ) = 0.
\end{equation}
Residual diffeomorphisms satisfy
\begin{equation}
(\nabla^{2}_x \xi_{\nu} + \nabla_{\nu} \nabla^{\mu} \xi_{\mu} - \xi_{\nu})=0
\end{equation}
and therefore $(\nabla^{2}_x -1) \nabla^{\mu} \xi_{\mu}  = 0$ allowing
us to further set $H^{(0)} = 4 \phi^{(0)}_3$.  To finally eliminate any
independent degrees of freedom of $H^{(0)}_{\mu \nu}$ consider
a solution to the homogeneous part of it's equations of motion.
It must be traceless, divergenceless, and satisfy
\begin{equation}
(\nabla^{2}_x + 2) H^{(0)}_{\mu \nu} =0
\end{equation}
from (\ref{Emunu}).
The remaining residual diffeomorphisms however satisfy
\begin{equation}
(\nabla^{2}_x + 2)\nabla_{(\mu}\xi_{\nu)} =0
\end{equation}
and $\nabla^{\mu} \xi_{\mu} = 0$, and can be used to eliminate 
$H^{(0)}_{\mu \nu}$ in terms of $\phi^{(0)}_3$.

Finally the gauge field $b^{(0)[12]}_{\mu}$ can be eliminated
by the $U(1)$ gauge symmetry exactly as described above.
A summary of the masses of the ${\cal N}=2$ supergravity scalars is
given in Table 1.  A final comment before leaving this section is
to note that for $k \geq 2$ half of the local degrees of freedom of
$B^{(k)}_{2 \mu}$ and all local degrees of freedom of 
$H^{(k)}_{\mu \nu}$ (in the $k \geq 1$ case as well) were eliminated
by equations of motion.  Normally these are eliminated by gauge
symmetries as in the $k=0$ case above.  In fact though their
elimination for $k \geq 2$ also follows by gauge symmetry.  
Four-dimensional diffeomorphism invariance must remove eight
``four''-dimensional fields.  Four of these fields were removed
immediately when we fixed the gauge (\ref{diffeofix}).  The
remaining four followed via the equations of motion.

\subsection{Completion of bosonic fields to ${\cal N} =8$ supergravity}

The remaining bosonic fields of ${\cal N}=8$ supergravity are the
scalars $W_{ABCD}$ and vectors $b_{\muhat}^{MN}$ for $M,N \neq 1,2$.
Before computing their mass spectrum let's first understand the $SU(2)
\times SU(6)$ transformation properties of the fields.  For the scalars
this follows trivially from their $SU(8)$ transformation properties.
For the vectors it is actually the field strength tensors (\ref{E7F})
that transform under $SU(8)$.  Substituting the vector and scalar
expansions (\ref{bmuexp},\ref{balphaexp},\ref{Wexp}) and the
expression for the bosonic part of $H_{\muhat \nuhat MN}$ 
in the symmetric gauge (\ref{Hbosonic}) we find
\begin{eqnarray}
{\cal F}^{(k)}_{MN} := \bigl(e^{\lambda \rho} \nabla_{\lambda} 
b_{\rho}^{(k)MN} + i k(k+1) b_{2}^{(k)MN} + i 2 (W^{(k)}_{12MN}
+ \bar{W}^{(k)12MN}) \bigr)
\label{calFMN}
\end{eqnarray}
where ${\cal F}_{\mu \nu MN} = - e_{\mu \nu}/\sqrt{2} \sum_k 
{\cal F}^{(k)}_{MN} Y_{(k)}$.  The fields ${\cal F}^{(k)}_{MN}$
therefore transform in the ${\bf 28}$ of $SU(8)$ and it's 
$SU(2) \times SU(6)$ transformation properties follow.  Note
that the scalar contribution to (\ref{calFMN}) vanishes
when either $M$ or $N$ is 1 or 2. 

To compute the remaining bosonic masses let's begin with the
$A=1,2;B,C,D \neq 1,2$ scalars $W_{ABCD}$ and the $M=1,2;N \neq 1,2$
vectors $g_{\muhat \nuhat}^{MN}$, which are already completely
decoupled from all other fields.  Substituting the harmonic expansions
(\ref{bmuexp},\ref{balphaexp},\ref{Wexp}) into the equations 
(\ref{V6},\ref{S20}) yields
\begin{mathletters}
\begin{eqnarray}
(\nabla^{2}_x - k(k+1)) W^{(k)}_{ABCD} & = & 0, \hspace{1cm} k \geq 0 \\
(2 \nabla^{\mu}\nabla_{[\mu} b_{\nu]}^{(k)MN} - k(k+1) b_{\nu}^{(k)MN})
& = & 0, \hspace{1cm} k \geq 0 \\
\nabla^{\mu} b_{\mu}^{(k)MN} & = & 0, \hspace{1cm} k \geq 1 \\
(\nabla^{2}_x - k(k+1)) b_{2}^{(k)MN} & = & 0, \hspace{1cm} k \geq 1.
\end{eqnarray}
\end{mathletters}
As before the two equations for the vector $b_{\mu}^{(k)MN}$ are equivalent
to
\begin{equation}
(\nabla^{2}_x - k(k+1)) e^{\lambda \rho} \nabla_{\lambda} b_{\rho}^{(k)MN}
=0, \hspace{1cm} k \geq 1.
\end{equation} 
Furthermore we can combine the equations for 
$e^{\lambda \rho} \nabla_{\lambda} b_{\rho}^{(k)MN}$ and $b_{2}^{(k)MN}$
into the $SU(8)$ covariant equation
\begin{equation}
(\nabla^{2}_x - k(k+1)) {\cal F}^{(k)}_{MN} = 0, \hspace{1cm} k \geq 1.
\end{equation}
The zero mode of the vector $b_{\mu}^{(0)MN}$ can be completely
eliminated as discussed before leaving us with two towers of complex
scalars of mass $m^2 = k(k+1)$ and degeneracy
$(2k+1)$ for
fixed group indices.  The $A=1,2;B,C,D \neq 1,2$ scalars $W^{(k)}_{ABCD}$ 
transform in the $({\bf 2},{\bf 20})$ of $SU(2) \times SU(6)$ and
the $M=1,2; N \neq 1,2$ scalars ${\cal F}^{(k)}_{MN}$ in the
$({\bf 2}, {\bf 6})$ representation.

Substituting the harmonic expansions (\ref{bmuexp},\ref{balphaexp},
\ref{Wexp}) into the remaining bosonic equations of motion
(\ref{V15},\ref{S15}) for the $M,N \neq 1,2$ vectors
$g_{\muhat \nuhat}^{MN}$ and the $C,D \neq 1,2$ scalars
$W_{12CD}$ we obtain
\begin{mathletters}
\begin{eqnarray}
\bigl((\nabla^{2}_x - k(k+1))W^{(k)}_{12CD} - 2 \bar{W}^{(k)12CD}
+ i (e^{\mu \nu} \nabla_{\mu} b_{\nu}^{(k)CD} + 
i k(k+1) b_{2}^{(k)CD}) \bigr) & = & 0, \hspace{1cm} k \geq 0 \\
\bigl( 2 \nabla^{\mu} \nabla_{[\mu}b_{\nu]}^{(k)MN} - k(k+1)
b_{\nu}^{(k)MN} + i 2 e_{\mu \nu} \nabla^{\mu}(W^{(k)}_{12MN}
- \bar{W}^{(k)12MN}) \bigr) & = & 0, \hspace{1cm} k \geq 0 \\
\nabla^{\mu} b_{\mu}^{(k)MN} & = & 0, \hspace{1cm} k \geq 1 \\
\bigl( (\nabla^{2}_x - k(k+1)) b_{2}^{(k)MN} - 2 (W^{(k)}_{12MN} +
\bar{W}^{(k)12MN}) \bigr) & = & 0, \hspace{1cm} k \geq 1.
\end{eqnarray}
\end{mathletters}
As before the equations for the vectors $b_{\mu}^{(k)MN}$ may
be rewritten as
\begin{equation}
\bigl( ( \nabla^{2}_x - k(k+1)) e^{\mu \nu} \nabla_{\mu} b_{\nu}^{(k)MN}
- i 2 \nabla^{2}_x ( W^{(k)}_{12MN} - \bar{W}^{(k)12MN}) \bigr) = 0,
\hspace{1cm} k \geq 1.
\end{equation}
After some rearranging the equations can be rewritten in $SU(2) \times
SU(6)$ covariant form as
\begin{mathletters}
\begin{eqnarray}
(\nabla^{2}_x - k(k+1) + 2) W^{(k)}_{12CD} + i {\cal F}^{(k)}_{CD} & = &
0, \hspace{1cm} k \geq 1 \\
(\nabla^{2}_x - k(k+1) -4) {\cal F}^{(k)}_{CD} - i 4(k(k+1)-2) 
W^{(k)}_{12CD} & = &
0, \hspace{1cm} k \geq 1.
\end{eqnarray}
\end{mathletters}
Diagonalizing the mass matrix produces two towers of complex scalars
with masses
\begin{equation}
m^2 = k(k-1), k^2+3k+2, \hspace{1cm} k \geq 1
\end{equation}
and with level $k$ degeneracy 
$(2k+1)$ for fixed group indices.  The zero mode
of the vector $b_{\mu}^{(0)MN}$ can be eliminated as described before
leaving us with one massive scalar to complete the
tower $m^2=k^2 + 3k +2$ to $k=0$.  Both towers of scalars transform
in the $({\bf 1},{\bf 15})$ of $SU(2) \times SU(6)$.

\section{Fermionic masses}
\label{section4}

\subsection{Spinor harmonic expansion on $S^2$}

A complete set of complex two-component spinors on $S^2$ is given
by $\psi_{slm}$ and $\rho_5 \psi_{slm}$ \cite{CamHig} ($\gamma$-matrix
conventions are given in the appendix) where $s=\pm$, $m=0,...,(l+1)$,
and
\begin{mathletters}
\begin{eqnarray}
(\rho^{\alpha} D_{\alpha} - i(l+1))\psi_{slm} & = & 0 \\
(\rho^{\alpha} D_{\alpha} + i(l+1))\rho_5 \psi_{slm} & = & 0.
\end{eqnarray}
\end{mathletters}
The spinors also satisfy the complex conjugation property
\begin{equation}
(\psi_{slm})^* = i (s) \rho_5 \psi_{-s,lm}.
\label{sbasisconj}
\end{equation} 
The 4-dimensional spinor and gravitino expansions therefore take the
form
\begin{mathletters}
\begin{eqnarray}
\lambda_{ABC} & = & \sum(\lambda^{(s,k)}_{+ABC} \otimes \psi_{(s,k)}
+ \lambda^{(s,k)}_{-ABC} \otimes \rho_5 \psi_{(s,k)} ) \label{lambdaexp} \\
\psi_{\mu A} & = & \sum(\psi^{(s,k)}_{+\mu A} \otimes \psi_{(s,k)}
+ \psi^{(s,k)}_{- \mu A} \otimes \rho_5 \psi_{(s,k)} ) \label{psimuexp} \\
\psi_{\alpha A} & = & \sum (\psi^{(s,k)}_{+A} \otimes D_{(\alpha)}\psi_{(s,k)}
+ \chi^{(s,k)}_{+A} \otimes \rho_{\alpha} \psi_{(s,k)}
+ \psi^{(s,k)}_{-A} \otimes D_{(\alpha)}(\rho_5 \psi_{(s,k)})
+ \chi^{(s,k)}_{-A} \otimes \rho_{\alpha} \rho_5 \psi_{(s,k)} )
\label{psialphaexp}
\end{eqnarray}
\end{mathletters}
where $D_{(\alpha)} := (D_{\alpha} - (1/2) \rho_{\alpha} \rho^{\beta}
D_{\beta})$ and we have again combined the $l,m$ indices into $k$.  The 
Majorana condition gives rise to 
\begin{equation}
(\psi^{(s,k)}_{+})^* = -i (s) \psi^{(-s,k)}_{-}
\end{equation}
for all $+/-$ fermionic coefficients in the above expansions. 

The linearized supersymmetry transformations for the gravitini
take the form
\begin{mathletters}
\begin{eqnarray}
\delta_s \psi_{\mu A} & = & (D_{\mu} \epsilon_A - \frac{i}{2}
\rho_{\mu} \otimes 1 \, \epsilon_{AB} \epsilon_B), \\
\delta_s \psi_{\alpha A} & = & (D_{\alpha} \epsilon_A - \frac{i}{2}
1 \otimes \rho_5 \rho_{\alpha} \epsilon_{AB} \epsilon_B)
\end{eqnarray}
\end{mathletters}
for $A =1,2$ and
\begin{equation}
\delta_s \psi_{\muhat A} = D_{\muhat} \epsilon_A, \hspace{1cm}
A \neq 1,2.
\end{equation}
Expanding $\epsilon_A$ as in (\ref{lambdaexp}) it is easy to see 
after using the identity $D_{\alpha}=D_{(\alpha)} + (1/2) \rho_{\alpha}
\rho^{\beta} D_{\beta}$ that $\chi^{(s,k)}_{\pm A}$ can be removed
for $k \geq 1, A=1,2$ and $k \geq 0, A \neq 1,2$.  The case $k=0,A=1,2$
is slightly more subtle because of the existence of the Killing
spinors $\psi_{(s,0)}$ which satisfy
\begin{equation}
D_{\alpha}\psi_{(s,0)} =
\frac{i}{2} \rho_{\alpha} \psi_{(s,0)}.
\end{equation}
The variations of $\chi^{(s,0)}_{\pm A}$ under supersymmetry
transformations for $A=1,2$ are easily shown to be
\begin{eqnarray}
\delta_s \chi^{(s,0)}_{+A} & = & \frac{i}{2} ( \epsilon^{(s,0)}_{+A} +
\epsilon_{AB} \epsilon^{(s,0)}_{-B}) \\
\delta _s \chi^{(s,0)}_{-A} & = & \frac{i}{2} 
\epsilon_{AB} ( \epsilon^{(s,0)}_{+B} +
\epsilon_{BC} \epsilon^{(s,0)}_{-C}) 
\end{eqnarray}
and therefore that $(\chi^{(s,0)}_{+A} + \epsilon_{AB} \chi^{(s,0)}_{-B})$ is
invariant and cannot be gauged away.
The opposite combination 
$(\chi^{(s,0)}_{+A} - \epsilon_{AB} \chi^{(s,0)}_{-B})$ however
can be gauged away.  The $\psi_{\alpha A}$ expansions then
simplify to
\begin{equation}
\psi_{\alpha A}  =  \sum_{k \geq 1} \bigl(\psi^{(s,k)}_{+A} 
\otimes D_{(\alpha)}\psi_{(s,k)}
+ \psi^{(s,k)}_{-A} \otimes D_{(\alpha)}(\rho_5 \psi_{(s,k)}) \bigr)
+ \bigl(\chi^{(s,0)}_{+A} \otimes \rho_{\alpha} \psi_{(s,0)} 
+ \epsilon_{AB} \chi^{(s,0)}_{+B} \otimes \rho_{\alpha} \rho_5 
\psi_{(s,0)} \bigr) 
\label{psialphaexp'}
\end{equation}
for $A=1,2$ and
\begin{equation}
\psi_{\alpha A}  =  \sum_{k \geq 1} \bigl(\psi^{(s,k)}_{+A} 
\otimes D_{(\alpha)}\psi_{(s,k)}
+ \psi^{(s,k)}_{-A} \otimes D_{(\alpha)}(\rho_5 \psi_{(s,k)}) \bigr)
\end{equation}
for $A \neq 1,2$.

\subsection{${\cal N} = 2$ fermi field content}

As in the bosonic case we again begin by finding the mass spectrum
for the ${\cal N} = 2$ fermi fields only.  This consists of 
the two gravitini $\psi_{\muhat A}$ for $A=1,2$.  Substituting the
expansions (\ref{psimuexp},\ref{psialphaexp'}) into the linearized
equations of motion (\ref{G1}) we obtain
\begin{mathletters}
\begin{eqnarray}
\bigl( \frac{i}{2}((k+1)^2 -1) \rho_{\mu} \psi^{(s,k)}_{+A} - (k+1)
\psi^{(s,k)}_{- \mu A} + \epsilon_{AB} \psi^{(s,k)}_{+ \mu B} \bigr) & = & 0 \\
\bigl( - \rho^{\mu} \psi^{(s,k)}_{+ \mu A} + \rho^{\mu} D_{\mu} 
\psi^{(s,k)}_{+A} - i \epsilon_{AB} \psi^{(s,k)}_{+B} \bigr) & = & 0 \\
\bigl( - \rho_4 e^{\mu \nu} D_{\mu} \psi^{(s,k)}_{- \nu A}
- \frac{i}{2} (k+1) \rho^{\mu} D_{\mu} \psi^{(s,k)}_{+A} 
- \frac{1}{2} (k+1) \epsilon_{AB} \psi^{(s,k)}_{+B} \bigr) & = & 0 \\
\end{eqnarray}
\end{mathletters}
for the $k \geq 1$ modes, the zero modes we handle separately.

Once again there are more equations than fields, however one equation
is not linearly independent.  The first equation, along with it's
complex conjugate, eliminates
$\psi^{(s,k)}_{\pm \mu A}$ in terms of $\psi^{(s,k)}_{\pm A}$.
After some trivial rearranging we find
\begin{equation}
(\rho^{\mu} D_{\mu} + (k+1)) \eta^{(s,k)}_{A} = 0
\end{equation}
where
\begin{equation}
\eta^{(s,k)}_{A} := (\psi^{(s,k)}_{+A} - i \psi^{(s,k)}_{-A})
\label{etadefn}
\end{equation}
Defining $AdS_2$ spinor masses by $m=|\kappa|$ for 
$(\rho^{\mu} D_{\mu} - \kappa) \lambda = 0$ we find two towers of complex
spinors, i.e., $A=1,2$, each with mass $m=(k+1)$ for $k \geq 1$
and degeneracy $2(k+1)$.  The two sets of spinors can be further
decomposed into two $({\bf 2}, {\bf 1})$ representations of 
$SU(2) \times SU(6)$.  
\vspace{5mm}

\noindent
$\bullet$ {\bf $k=0$ sector} 
\vspace{5mm}

The equations of motion in the $l=0$ sector are
\begin{mathletters}
\begin{eqnarray}
\bigl(i 2 \rho_4 e^{\mu \nu} D_{\nu} \chi^{(s,0)}_{+A} + \rho^{\mu}
\epsilon_{AB} \chi^{(s,0)}_{+B} + e^{\mu \nu} \rho_4 (\psi^{(s,0)}_{+ \nu A}
+ \epsilon_{AB} \psi^{(s,0)}_{- \nu B}) \bigr) & = & 0 \\
\bigl(\rho^{\mu} D_{\mu} \chi^{(s,0)}_{+A} - i \epsilon_{AB}
\chi^{(s,0)}_{+B} - \rho_4 e^{\mu \nu} D_{\mu} \psi^{(s,0)}_{- \nu A}
- \frac{i}{2} \rho^{\mu} \psi^{(s,0)}_{+ \mu A} \bigr) & = & 0.
\label{2ndpsil=0}
\end{eqnarray}
\end{mathletters}
The first equation along with it's complex conjugate are enough
to show that
\begin{equation}
(\rho^{\mu} D_{\mu} + 1)(\delta_{AB} + i \epsilon_{AB})\chi^{(s,0)}_{+B} =0
\end{equation}
and
\begin{equation}
\psi^{(s,0)}_{+ \mu A} = - \epsilon_{AB} \psi^{(s,0)}_{- \mu B}.
\end{equation}
Therefore we have a single complex spinor with mass $m=1$ transforming
under the $({\bf 2},{\bf 1})$ representation of $SU(2) \times SU(6)$.
We also have
the two-dimensional gravitino $\psi^{(s,0)}_{+ \mu A}$.  This field
however can be gauged away by using the residual supersymmetry transformations
satisfying $\epsilon^{(s,0)}_{+A} = - \epsilon_{AB} \epsilon^{(s,0)}_{-B}$.
Specifically one can first demand
\begin{equation}
\rho^{\mu} \psi^{(s,0)}_{+ \mu A} = - 4 \epsilon_{AB} \psi^{(s,0)}_{+B}.
\end{equation}
The equation of motion (\ref{2ndpsil=0}) then reduces to 
$e^{\mu \nu}D_{\mu} \psi^{(s,0)}_{+ \nu A}=0$, which is also
the equation satisfied by residual supersymmetry transformations
$\delta_s \psi^{(s,0)}_{+ \mu A}$ so that $\psi^{(s,0)}_{+ \mu A}$
can be removed locally.

\subsection{Completion of the fermionic fields to ${\cal N}=8$}

The remaining fermionic fields are the spinors $\lambda_{ABC}$ and
the $A \neq 1,2$ gravitini $\psi_{\muhat A}$.  The $A =1,2; B,C \neq 1,2$
spinors are already completely decoupled so their masses follow
almost immediately.  Substituting the expansion (\ref{lambdaexp})
into the equation of motion (\ref{Sp15}) we find after taking
appropriate linear combinations
\begin{equation}
\bigl(\rho^{\mu} D_{\mu} - (k+1) \bigr) \xi^{(s,k)}_{ABC} =0 
\end{equation}
where 
\begin{equation}
\xi^{(s,k)}_{ABC} := (\lambda^{(s,k)}_{+ABC} + i \lambda^{(s,k)}_{-ABC}).
\label{xidefn}
\end{equation}
Therefore we have one tower of complex spinors with mass $m=(k+1)$, 
degeneracy
$2(k+1)$ at level $k \geq 0$, and transforming in the
$({\bf 2},{\bf 15})$ of $SU(2) \times SU(6)$.

The $A,B,C \neq 1,2$ spinor masses are also straightforward to compute.
Substituting the expansion (\ref{lambdaexp}) as well as the background
field strength (\ref{Gbackground}) into the equation of motion
(\ref{Sp20})
yields after some rearranging\footnote{These expressions may appear
incorrect at first sight because of the mixing of up and down
$SU(8)$ indices.  The $\xi^{(s,k)}_{ABC}$ fields however do not
transform covariantly under $SU(8)$ because of the $\gamma_5$ matrix
appearing in the transformation law of $\lambda_{ABC}$.  Their transformation
law is easy to find and indeed one may show that an $SU(6)$ transformation
maps these equations into themselves and their complex conjugates (after
letting $s \rightarrow -s$).}
\begin{mathletters}
\begin{eqnarray}
\bigl(\rho^{\mu} D_{\mu} - k \bigr) \bigl(\xi^{(s,k)}_{ABC} + i \frac{\eta}{6}
\epsilon^{12ABCFGH} \xi^{(s,k)}_{FGH} \bigr) & = & 0 \\
\bigl(\rho^{\mu} D_{\mu} - (k+2) \bigr) \bigl(\xi^{(s,k)}_{ABC} - 
i \frac{\eta}{6}
\epsilon^{12ABCFGH} \xi^{(s,k)}_{FGH} \bigr) & = & 0.
\end{eqnarray}
\end{mathletters}
We therefore have two towers of complex spinors with masses $m=k$ and
$m=(k+2)$ respectively.  Both towers have degeneracy
$2(k+1)$ at
level $k \geq 0$ and transform in the $({\bf 1},{\bf 20})$ of
$SU(2) \times SU(6)$.

The remaining coupled fermions are the  $\lambda_{12A}$ spinors
and the $A \neq 1,2$ gravitini $\psi_{\muhat A}$.  Substituting
the expansions (\ref{lambdaexp},\ref{psimuexp},\ref{psialphaexp})
into the equations of motion (\ref{Sp6},\ref{G6}) gives rise
to
\begin{mathletters}
\begin{eqnarray}
\bigl( \rho^{\mu} D_{\mu} \lambda^{(s,k)}_{+12A} - i(k+1) 
\lambda^{(s,k)}_{-12A} - \frac{i}{\sqrt{2}} \rho^{\mu} 
\psi^{(s,k)}_{- \mu C} \bigr) & = & 0 \label{lambda12C} \\
\bigl( \frac{i}{2} ((k+1)^2 -1) \rho_{\mu} \psi^{(s,k)}_{+A}
- (k+1) \psi^{(s,k)}_{- \mu A} + \frac{1}{\sqrt{2}} \rho_{\mu}
\lambda^{(s,k)}_{-12A} \bigr) & = & 0 \label{psimu6} \\
\bigl(\rho^{\mu} D_{\mu} \psi^{(s,k)}_{+A} - \rho^{\mu} 
\psi^{(s,k)}_{+ \mu A} \bigr) & = & 0 \label{psialpha1} \\
\bigl(- \rho_4 e^{\mu \nu} D_{\mu} \psi^{(s,k)}_{- \nu A}
- \frac{i}{2} (k+1) \rho^{\mu} D_{\mu} \psi^{(s,k)}_{+A}
+ \frac{i}{\sqrt{2}} \lambda^{(s,k)}_{+12A} \bigr) & = & 0
\label{psialpha2}
\end{eqnarray}
\end{mathletters}
for $k \geq 1$.
The second equation eliminates $\psi^{(s,k)}_{- \mu A}$ in terms
of the other fields (and similarly $\psi^{(s,k)}_{+ \mu A}$ after
complex conjugation).  One of the remaining equations is redundant,
while the others imply after taking appropriate linear combinations
\begin{mathletters}
\begin{eqnarray}
\bigl(\rho^{\mu} D_{\mu} - k \bigl) \bigr(\eta^{(s,k)}_{A} +
\frac{\sqrt2}{k+2} \xi^{(s,k)}_{12A} \bigl) & = & 0 \\
\bigl(\rho^{\mu} D_{\mu} - (k+2) \bigl) \bigr(\eta^{(s,k)}_{A} -
\frac{\sqrt2}{k} \xi^{(s,k)}_{12A} \bigl) & = & 0 
\end{eqnarray}
\end{mathletters}
where $\xi^{(s,k)}_{ABC}$ is defined in (\ref{xidefn}) and 
$\eta^{(s,k)}_{A}$ in (\ref{etadefn}).

For the zero modes the equations are given by
(\ref{lambda12C},\ref{psimu6}) with
$k=0$ and $\psi^{(s,0)}_{\pm A} =0$.  
It follows that
$\psi^{(s,k)}_{- \mu A}$ can be eliminated and that
\begin{equation}
\bigl( \rho^{\mu} D_{\mu} - 2 \bigr) \xi^{(s,0)}_{12A} = 0.
\end{equation}
For these fields we therefore find two towers of complex spinors, one with
mass $m=k$ for $k \geq 1$ and the other with mass $m=(k+2)$ for
$k \geq 0$.  Both have degeneracy $2(k+1)$ at level $k$ and transform
in the $({\bf 1},{\bf 6})$ of $SU(2) \times SU(6)$.
The complete set of fermion masses are summarized in Table 2.

\section{$SU(1,1|2)$ multiplets}
\label{section5}

Irreducible representations of $SU(1,1|2)$ are labelled by the
eigenvalues of the $L_0$ and $J_0$ generators of the
$SL(2,R) \times SU(2)$ subalgebra.  The irreducible representations
have been constructed in \cite{Gun2} and in particular the
so-called short multiplets which will be of interest here.  The short
multiplet
irreducible representations consist of the states
\begin{equation}
D^{(k)}(k) \oplus 2 D^{(k-1/2)}(k+1/2) \oplus D^{(k-1)}(k+1)
\label{kirrep}
\end{equation}
for half-integer $k \geq 1/2$ (where in the $k=1/2$ case it is
understood that the $D^{(-1/2)}(3/2)$ states are missing).  We now proceed
to show that the states listed in Tables 1 and 2 fill various
short multiplet representations of $SU(1,1|2)$.

An irreducible representation of $SU(2)$ labelled by $J_0$
eigenvalue $j$ has $(2j+1)$ states.  For the harmonic expansions
of the scalar fields (with degeneracy $(2k+1)$) this implies
$j=k$.  For the spinors with degeneracy $2(k+1)$ we
have $j=(k+1/2)$.  The $L_0$ eigenvalue of the states comes from
the $AdS/CFT$ map developed in \cite{GubKlePol,Wit}.  The conformal
weight of a boundary conformal field corresponding to an $AdS_2$
scalar was shown to be
\begin{equation}
h_{scalar} = \frac{1}{2} (1 + \sqrt{1 + 4 m^2}),
\label{scalarcw}
\end{equation}
and for a spinor \cite{Sfe} 
\begin{equation}
h_{spinor} = m + \frac{1}{2}.
\end{equation}
Plugging in the various scalar and spinor masses that we have found
results in the conformal weights shown in Tables 1 and 2 respectively.

The ${\cal N}=8$ supergravity fields also carry
$SU(2) \times SU(6)$ indices, and therefore so will the
boundary states.  We therefore label the complete boundary states as
\begin{equation}
D^{(j)}(h)({\bf R_2} \times {\bf R_6})
\end{equation}
where as above $h$ and $j$ label the $SL(2,R) \times SU(2)$ content
of $SU(1,1|2)$ and ${\bf R_2} \times {\bf R_6}$ labels the 
$SU(2) \times SU(6)$
content.  From Tables 1 and 2 we can now read off the representations.
For the ${\cal N} = 2$ sector we find two sets of
\begin{equation}
D^{(k)}(k)({\bf 1},{\bf 1}) \oplus D^{(k-1/2)}(k+1/2)({\bf 2},{\bf 1})
\oplus D^{(k-1)}(k+1)({\bf 1},{\bf 1})
\end{equation}
for $k \geq 1$ and $k \geq 2$.  For the remaining fields we find the
representations
\begin{eqnarray}
D^{(k+1/2)}(k+1/2)({\bf 1},{\bf 20}) \oplus D^{(k)}(k+1)({\bf 2},{\bf 20})
\oplus D^{(k-1/2)}(k+3/2)({\bf 1},{\bf 20}) \hspace{5mm} k & \geq & 0 \\
D^{(k+1/2)}(k+1/2)({\bf 1},{\bf 6}) \oplus D^{(k)}(k+1)({\bf 2},{\bf 6})
\oplus D^{(k-1/2)}(k+3/2)({\bf 1},{\bf 6}) \hspace{5mm} k & \geq & 1 \\
D^{(k)}(k)({\bf 1},{\bf 15}) \oplus D^{(k-1/2)}(k+1/2)({\bf 2},{\bf 15})
\oplus D^{(k-1)}(k+1)({\bf 1},{\bf 15}) \hspace{5mm} k & \geq & 1.
\end{eqnarray}

\section{Conclusions}
\label{section6}

We have computed the mass spectrum of the ${\cal N}=8$
supergravity theory about $AdS_2 \times S^2$.  We have
shown that the corresponding spectrum of states in 
the boundary conformal field theory lies in short multiplets
of the $AdS_2$ supergroup $SU(1,1|2)$.  The states further
carry $SU(2) \times SU(6)$ indices inherited from the
spontaneously broken $SU(8)$ gauge group of the supergravity
theory.

Eventually one hopes to identify these boundary states
in the $d=1$, ${\cal N} = 4$ $n$-particle Calogero model, conjectured
\cite{GibTow}
to be the dual conformal quantum mechanics model to the
the near horizon region of the
intersecting D3-brane solution \cite{D3soln}.  As noted in
\cite{GibTow} though, this theory has yet to be constructed.
Recently however the one-dimensional single particle
Calogero model has
been constructed \cite{AkuKud} for an arbitrary number of supersymmetries.
Generalization of the model to arbitrary $n$ should be straightforward.
Once this is done a detailed check on the correspondence between the two 
theories should be possible.

We have ignored the boundary degrees of freedom except for one
massless vector in the ${\cal N}=2$ sector which was needed to
fill an $SU(1,1|2)$ representation.  Such fields may also be
necessary in realizing the $AdS/CFT$ duality.
It should be straightforward to extract this information from
the results in this paper.

While this work was being written up another paper \cite{Mic} 
appeared reporting similar results for the ${\cal N}=2$ sector
of the ${\cal N}=8$ supergravity theory discussed here.
\cite{Mic} also considered in detail the boundary degrees of
freedom for ${\cal N}=2$ supergravity.

\section*{Acknowledgements}
This work was supported by the Natural Sciences and Engineering
Research Council of Canada at the University of Alberta.

\appendix
\section*{Gamma Matrices}
\label{appendix}

Our gamma matrix and spinor conventions are
\begin{eqnarray}
\begin{array}{cc}
\mbox{$\{\gamma^a,\gamma^b \}  =  2 \eta^{ab}$}, &
\mbox{$(\gamma^a)^{\dagger} = \gamma^0 \gamma^a \gamma^0$}  \\
\mbox{$(\gamma^a)^a = \gamma^a$}, & 
\mbox{$(\gamma^a)^T = \gamma^0 \gamma^a \gamma^0$} \\
\mbox{$\gamma_5 = i \gamma^0 \gamma^1 \gamma^2 \gamma^3$}, &
\mbox{$\{\gamma_5, \gamma^a \} = 0$} \\
\mbox{$\gamma^{\mu} = e^{\mu}_{\;\;\;a} \gamma^a$}, & 
\mbox{$\bar{\lambda} = 
\lambda^{\dagger} \gamma^0$} \\
\mbox{$\gamma^{[a_1 a_2 \cdots a_n ]} = \gamma^{[ a_1} \gamma^{a_2}
\cdots \gamma^{a_n ]}$} &
\end{array}
\end{eqnarray}
where the metric signature is $(-,+,+,+)$ and $a$ and $\mu$ both take
values in $0,1,2,3$.
By choosing the gamma matrices to be real, the Majorana condition
reduces to $\lambda^* = \lambda$.  We split the $\gamma^a$-matrices
under $SO(1,3) \rightarrow SO(1,1) \otimes SO(2)$ as
\begin{eqnarray}
\gamma^a = \rho^a \otimes \rho_5, \hspace{1cm} \gamma^i = 1 \otimes \rho^i
\end{eqnarray}
where now $a = 0,1$ and $i=2,3$.  We define the $SO(1,1)$ and $SO(2)$
``$\gamma_5$'' matrices by $\rho_4 := \rho^0 \rho^1$ and $\rho_5 :=
i \rho^2 \rho^3$ respectively.  One possible choice of $\rho$-matrices
satisfying the above $\gamma$-matrix conventions is given by
\begin{eqnarray}
\begin{array}{cc}
\mbox{$\rho^0$}  =  \left( \begin{array}{cc} 0 & \mbox{$i$} \\
\mbox{$i$} & 0 \end{array} \right), &
\mbox{$\rho^1$}  =  \left( \begin{array}{cc} 0 & \mbox{$-i$} \\
\mbox{$i$} & 0 \end{array} \right) \\
\\
\mbox{$\rho^2$}  =  \left( \begin{array}{cc} 0 & 1 \\
1 & 0 \end{array} \right), &
\mbox{$\rho^3$}  =  \left( \begin{array}{cc} 1 & 0 \\
0 & -1 \end{array} \right)
\end{array}
\end{eqnarray}

\newpage

\begin{picture}(410,270)(0,-20)
\put(0,210){\makebox(410,40){Table 1 - Scalar Fields}}

\put(0,180){\framebox(110,30){Fields}}
\put(110,180){\framebox(80,30){$SU(2) \times SU(6)$}}
\put(190,180){\framebox(40,30){modes}}
\put(230,180){\framebox(60,30){mass}}
\put(290,180){\framebox(60,30){\shortstack{conformal\\dimension}}}
\put(350,180){\framebox(60,30){degeneracy}}

\put(0,140){\framebox(110,40){\shortstack{$\phi^{(k)}_3$\\$b^{(k)}_2$}}}
\put(110,160){\framebox(80,20){$({\bf 1},{\bf 1})$}}
\put(110,140){\framebox(80,20){$({\bf 1},{\bf 1})$}}
\put(190,160){\framebox(40,20){$k \geq 1$}}
\put(190,140){\framebox(40,20){$k \geq 0$}}
\put(230,160){\framebox(60,20){$k(k-1)$}}
\put(230,140){\framebox(60,20){$k^2 + 3k + 2$}}
\put(290,160){\framebox(60,20){$k$}}
\put(290,140){\framebox(60,20){$k+2$}}
\put(350,160){\framebox(60,20){$2k+1$}}
\put(350,140){\framebox(60,20){$2k+1$}}

\put(0,100){\framebox(110,40)
{\shortstack{$e^{\mu \nu} \nabla_{\mu} b^{(k)[12]}_{\nu}$\\
$e^{\mu \nu} \nabla_{\mu} B^{(k)}_{2 \nu}$}}}
\put(110,120){\framebox(80,20){$({\bf 1},{\bf 1})$}}
\put(110,100){\framebox(80,20){$({\bf 1},{\bf 1})$}}
\put(190,120){\framebox(40,20){$k \geq 2$}}
\put(190,100){\framebox(40,20){$k \geq 1$}}
\put(230,120){\framebox(60,20){$k(k-1)$}}
\put(230,100){\framebox(60,20){$k^2 + 3k + 2$}}
\put(290,120){\framebox(60,20){$k$}}
\put(290,100){\framebox(60,20){$k+2$}}
\put(350,120){\framebox(60,20){$2k+1$}}
\put(350,100){\framebox(60,20){$2k+1$}}

\put(0,60){\framebox(110,40){\shortstack{$W_{ABCD}$\\
$A = 1,2; B,C,D \neq 1,2$}}}
\put(110,60){\framebox(80,40){$({\bf 2},{\bf 20})$}}
\put(190,60){\framebox(40,40){$k \geq 0$}}
\put(230,60){\framebox(60,40){$k(k+1)$}}
\put(290,60){\framebox(60,40){$k+1$}}
\put(350,60){\framebox(60,40){$2k+1$}}

\put(0,20){\framebox(110,40){\shortstack{${\cal F}^{(k)}_{MN}$\\
$M = 1,2; N \neq 1,2$}}}
\put(110,20){\framebox(80,40){$({\bf 2},{\bf 6})$}}
\put(190,20){\framebox(40,40){$k \geq 1$}}
\put(230,20){\framebox(60,40){$k(k+1)$}}
\put(290,20){\framebox(60,40){$k+1$}}
\put(350,20){\framebox(60,40){$2k+1$}}



\put(0,-20){\framebox(110,40)
{\shortstack{$W^{(k)}_{12CD}; C,D \neq 1,2$\\
${\cal F}^{(k)}_{CD}; C,D \neq 1,2$}}}
\put(110,0){\framebox(80,20){$({\bf 1},{\bf 15})$}}
\put(110,-20){\framebox(80,20){$({\bf 1},{\bf 15})$}}
\put(190,0){\framebox(40,20){$k \geq 1$}}
\put(190,-20){\framebox(40,20){$k \geq 0$}}
\put(230,0){\framebox(60,20){$k(k-1)$}}
\put(230,-20){\framebox(60,20){$k^2 + 3k + 2$}}
\put(290,0){\framebox(60,20){$k$}}
\put(290,-20){\framebox(60,20){$k+2$}}
\put(350,0){\framebox(60,20){$2k+1$}}
\put(350,-20){\framebox(60,20){$2k+1$}}
\end{picture}

\vspace{1cm}

\begin{picture}(410,230)(0,20)
\put(0,210){\makebox(410,40){Table 2 - Spinor Fields}}

\put(0,180){\framebox(110,30){Fields}}
\put(110,180){\framebox(80,30){$SU(2) \times SU(6)$}}
\put(190,180){\framebox(40,30){modes}}
\put(230,180){\framebox(60,30){mass}}
\put(290,180){\framebox(60,30){\shortstack{conformal\\dimension}}}
\put(350,180){\framebox(60,30){degeneracy}}

\put(0,140){\framebox(110,40){\shortstack{$\eta^{(s,k)}_{A}$\\$A=1,2$}}}
\put(110,160){\framebox(80,20){$({\bf 2},{\bf 1})$}}
\put(110,140){\framebox(80,20){$({\bf 2},{\bf 1})$}}
\put(190,160){\framebox(40,20){$k \geq 1$}}
\put(190,140){\framebox(40,20){$k \geq 0$}}
\put(230,160){\framebox(60,20){$k+1$}}
\put(230,140){\framebox(60,20){$k+1$}}
\put(290,160){\framebox(60,20){$k+\frac{3}{2}$}}
\put(290,140){\framebox(60,20){$k+\frac{3}{2}$}}
\put(350,160){\framebox(60,20){$2(k+1)$}}
\put(350,140){\framebox(60,20){$2(k+1)$}}

\put(0,100){\framebox(110,40)
{\shortstack{$ \xi^{(s,k)}_{ABC}$\\
$A=1,2; B,C \neq 1,2$}}}
\put(110,100){\framebox(80,40){$({\bf 2},{\bf 15})$}}
\put(190,100){\framebox(40,40){$k \geq 0$}}
\put(230,100){\framebox(60,40){$k+1$}}
\put(290,100){\framebox(60,40){$k+\frac{3}{2}$}}
\put(350,100){\framebox(60,40){$2(k+1)$}}

\put(0,60){\framebox(110,40){\shortstack{$\xi^{(s,k)}_{ABC}$\\
$A,B,C \neq 1,2$}}}
\put(110,80){\framebox(80,20){$({\bf 1},{\bf 20})$}}
\put(110,60){\framebox(80,20){$({\bf 1},{\bf 20})$}}
\put(190,80){\framebox(40,20){$k \geq 0$}}
\put(190,60){\framebox(40,20){$k \geq 0$}}
\put(230,80){\framebox(60,20){$k$}}
\put(230,60){\framebox(60,20){$k+2$}}
\put(290,80){\framebox(60,20){$k+\frac{1}{2}$}}
\put(290,60){\framebox(60,20){$k+\frac{5}{2}$}}
\put(350,80){\framebox(60,20){$2(k+1)$}}
\put(350,60){\framebox(60,20){$2(k+1)$}}

\put(0,20){\framebox(110,40)
{\shortstack{$\xi^{(s,k)}_{12C}; C \neq 1,2$\\
$\eta^{(s,k)}_{C}; C \neq 1,2$}}}
\put(110,40){\framebox(80,20){$({\bf 1},{\bf 6})$}}
\put(110,20){\framebox(80,20){$({\bf 1},{\bf 6})$}}
\put(190,40){\framebox(40,20){$k \geq 1$}}
\put(190,20){\framebox(40,20){$k \geq 0$}}
\put(230,40){\framebox(60,20){$k$}}
\put(230,20){\framebox(60,20){$k+2$}}
\put(290,40){\framebox(60,20){$k+\frac{1}{2}$}}
\put(290,20){\framebox(60,20){$k+\frac{5}{2}$}}
\put(350,40){\framebox(60,20){$2(k+1)$}}
\put(350,20){\framebox(60,20){$2(k+1)$}}
\end{picture}
 
\end{document}